\documentclass[twocolumn,twoside]{IEEEtran}
\usepackage{mathpple}
\usepackage{times}

\usepackage{amsmath}  % Define \boldsymbol (in amsbsy too) and align
\usepackage{amssymb}  % Define \mathbb (in amsfonts too)
\usepackage{mathrsfs} % Define \mathscr, a script font

\usepackage{theorem}  % Helps in rendering theorems, etc.
\usepackage{cite}     % Gives multiple references as intervals
\usepackage{comment}  % Comments out with \begin{comment} ... \end{comment}

\usepackage{upref}
\usepackage{amsfonts}

\usepackage{verbatim}

\usepackage[dvipsnames,usenames]{color}
\usepackage{graphicx}

\newcommand{\be}[1]{\begin{equation}\label{#1}}
\newcommand{\ee}{\end{equation}}

\newcommand{\bc}{\begin{center}}
\newcommand{\ec}{\end{center}}

\renewcommand{\le}{\leqslant}
\renewcommand{\leq}{\leqslant}
\renewcommand{\ge}{\geqslant}
\renewcommand{\geq}{\geqslant}

\newcommand{\Cref}[1]{Co\-rol\-la\-ry\,\ref{#1}}

\theoremstyle{plain} \theorembodyfont{\normalfont\slshape}

\newtheorem{thm}{Theorem$\!$}
\newenvironment{theorem}{\begin{thm}\hspace*{-1ex}{\bf.}}{\end{thm}}

\newtheorem{prop}[thm]{Proposition$\!$}

\newtheorem{lem}[thm]{Lemma$\!$}

\newtheorem{cor}[thm]{Corollary$\!$}
\newenvironment{corollary}{\begin{cor}\hspace*{-1ex}{\bf.}}{\end{cor}}

\newtheorem{defi}{Definition.}

\theorembodyfont{\normalfont}

\newtheorem{exam}{Example$\!$}
\newenvironment{example}{\begin{exam}\hspace*{-1ex}{\bf .}}{\end{exam}}

\newtheorem{remrk}{Remark$\!$}

\definecolor{Codecolor}{named}{White}  %{Tan}
\newcommand{\Copen}{\mbox{\{\kern-5.50pt\{}}
\newcommand{\Cclose}{\mbox{\}\kern-5.50pt\}}}
\newcommand{\Cslash}{\mbox{$\backslash\kern-6.02pt\backslash$}}

\def\beginofproof{\noindent {\bf Proof. }}
\def\endofproof{\hfill$\Box$}

\begin{document}

% paper title
\title{Perspectives on Balanced Sequences}

%\author{\large Jos H. Weber,~\IEEEmembership{Senior Member,~IEEE,}, Kees~A.~Schouhamer Immink,~\IEEEmembership{Fellow,~IEEE}, %Paul~H.~Siegel,~\IEEEmembership{Fellow,~IEEE}, and Theo~G.~Swart,~\IEEEmembership{Member,~IEEE}
\author{\large Jos H. Weber, Kees~A.~Schouhamer Immink, Paul~H.~Siegel, and Theo~G.~Swart
\thanks{Submission to IEEE Transactions on Information Theory,  September 28, 2012.}
\thanks{J. H. Weber is with Delft University of Technology, Delft, The Netherlands
(e-mail: j.h.weber@tudelft.nl).
K.A.~Schouhamer~Immink is with Turing Machines Inc., Rotterdam, The Netherlands
(e-mail: immink@turing-machines.com).
P.H.\ Siegel is with the Department of Electrical and Computer Engineering and the Center for Magnetic Recording Research, University of California, San Diego, La Jolla, CA 92093, U.S.A. (e-mail:    psiegel@ucsd.edu).
T.G.~Swart is with the Department of Electrical and Electronic Engineering Science, University of Johannesburg, Auckland Park, South Africa
(e-mail: tgswart@uj.ac.za).}
\thanks{P.H. Siegel was supported in part by the Center for Magnetic Recording Research at the University of California, San Diego. Portions of this work were done while J.H.~Weber and P.H.~Siegel were visiting the Centre Interfacultaire Bernoulli (CIB) at the Ecole Polytechnique F\'{e}d\'{e}rale de Lausanne (EPFL) in Lausanne, Switzerland, in the autumn of 2011. Part of this work was presented at IEEE International Symposium on Information Theory, Seoul, Korea, 2009.} \vspace{-2ex}}

%\author{\large Eitan~Yaakobi,~\IEEEmembership{Student Member,~IEEE,} Paul~H.~Siegel,~\IEEEmembership{Fellow,~IEEE,} Alexander~Vardy,~\IEEEmembership{Fellow,~IEEE,} and Jack K. Wolf,~\IEEEmembership{Life Fellow,~IEEE}
%\thanks{Eitan Yaakobi, Paul H.\ Siegel, Alexander Vardy and Jack K.\ Wolf are with the Department of Electrical and Computer Engineering and the Center for Magnetic Recording Research, University of California at San Diego, La Jolla, CA 92093, U.S.A. (e-mail: \{eyaakobi,psiegel,avardy,jwolf\}@ucsd.edu).}
%\thanks{This research was supported in part by the University of California Lab Fees Research Program, Award No. 09-LR-06-118620-SIEP and the Center for Magnetic Recording Research at the University of California, San Diego. Part of the results in the paper were presented at the IEEE International Symposium on Information Theory, Austin, Texas, June 12 to June 18, 2010 (reference~\cite{YSVW10}).}\vspace{-2ex}}

\maketitle

\bibliographystyle{IEEEtranS}
%\vspace{6ex}
\begin{abstract}
We examine and compare several different classes of ``balanced'' block codes over $q$-ary alphabets, namely \emph{symbol-balanced} (SB) codes, \emph{charge-balanced} (CB) codes, and \emph{polarity-balanced} (PB) codes.  Known results on the maximum size and asymptotic minimal redundancy of SB and CB codes are reviewed. We then determine the maximum size and asymptotic minimal redundancy of PB codes and of codes which are both CB and PB. We also propose efficient Knuth-like encoders and decoders for all these types of balanced codes.
\end{abstract}

\section{Introduction}
\label{intro}

\noindent
There are several different classes of block codes over a $q$-ary integer alphabet that can be described as being ``balanced'' in some sense. Consider, for example, the symmetric alphabets ${\cal A}_q=\{-q+1,-q+3,-q+5,\ldots,q-3, q-1\}$ that arise in the context of pulse amplitude modulation (PAM), e.g., ${\cal A}_4=\{-3,-1,+1,+3\}$, ${\cal A}_5=\{-4,-2,0,+2,+4\}$. We say that a code is \emph{symbol-balanced} (SB) over ${\cal A}_q$ if, in each codeword, all $q$ alphabet symbols appear equally often. A \emph{charge-balanced} (CB) code is one in which the sum of the symbols in each codeword is zero. We also define \emph{polarity-balanced} (PB) codes, for which, in every codeword, the number of positive symbols equals the number of negative symbols. For $q$ odd, this definition does not constrain the number of zero symbols.

\begin{figure}
\centering%
\includegraphics[width=.9\columnwidth]{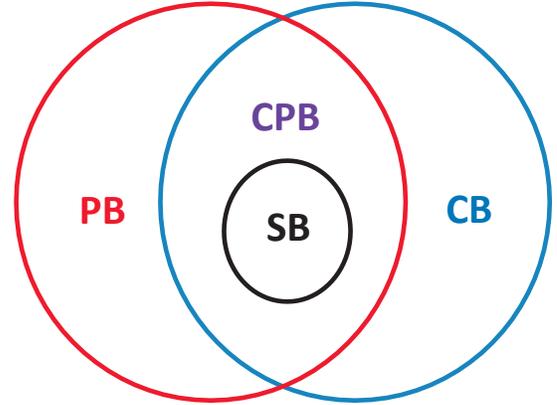}
\caption{Relationships among the symbol-balanced (SB), charge-balanced (CB), polarity-balanced (PB), and charge $\&$ polarity-balanced (CPB) properties.}
\label{bcfig}
\end{figure}

It is easy to see that for $q=2$, i.e., for bipolar sequences of even length $n$, these three notions of being ``balanced'' are completely equivalent. For $q=3$, i.e., for sequences over the alphabet $\{-2,0,+2\}$, the notions of CB and PB are equivalent, but the SB sequences form a proper subset of the set of CB and PB sequences. For example, the sequence $(-2,-2,+2,0,-2,+2,+2,+2,-2)$ of length 9 is CB and PB, but not SB. For $q\ge 4$, all three notions are mutually distinct. Any sequence which is SB is also CB and PB, but there do exist sequences which are PB but not CB (e.g., $(-3,-1,+1,+1)$ over ${\cal A}_4$) and sequences which are CB but not PB (e.g., $(+3,-1,-1,-1)$ over ${\cal A}_4$). Furthermore, there exist sequences which are both CB and PB (denoted as CPB) but not SB (e.g., $(-3,-3,+3,+3)$ over ${\cal A}_4$). In conclusion, the general relationship among the balancing criteria discussed above can be represented by the Venn diagram shown in Fig.~\ref{bcfig}.

Balanced codes have found applications in digital communications and data storage technology \cite{I2004}.
They have been widely studied in the literature, particularly for the binary case, e.g., \cite{AB1990}, \cite{AB1994}, \cite{ABCO1988}, \cite{K1986}, \cite{TB1999}, \cite{TCB1996}. Some constructions also take into account error correction capabilities, e.g., \cite{AB1993}, \cite{MRV2009}, \cite{TB1989}, \cite{WIF2012}. Results for non-binary alphabets have been presented for the SB and CB cases,
albeit under different (or no specific) names, e.g., \cite{MT2006} (SB) and \cite{CGV1991}, \cite{TV1999} (CB). To the best of our knowledge, the PB concept for non-binary sequences is new and has not been studied before. It is of particular interest for applications which demand a balancing of positive and negative symbols, possibly in combination with a charge constraint. In this paper, we determine the number of $q$-ary PB sequences of length $n$ as well as the number of $q$-ary sequences of length $n$ which are CPB, i.e., both CB and PB. From this, we derive expressions for the minimum redundancy of PB and CPB codes, which are compared to the corresponding expressions for SB and CB codes.

A celebrated method to generate and decode bipolar balanced sequences of even length $n$ was presented by Knuth \cite{K1986}. The key idea is to invert the first $z$ symbols of the information sequence such that the resulting sequence is balanced. Knuth showed that it is always possible to find at least one such balancing index $z$. By communicating the value of $z$ through a (balanced) prefix, decoding can be performed by inverting the first $z$ symbols of the coded sequence. The redundancy of this elegant method is roughly $\log_2(n)$, which is about twice the minimum and can thus be considered as a price to be paid for simplicity. In this paper, we extend Knuth's method, which assumes bipolar sequences, to larger alphabets. In particular, we present Knuth-like design methods for all balancing perspectives under consideration, i.e., for SB, CB, PB, and CPB.

The rest of this paper is organized as follows. In Section~\ref{pre}, some definitions and preliminaries are presented. Then,
in Section~\ref{redundancy}, we first review known expressions for the maximum sizes of $q$-ary SB and CB codes of length $n$, as well as the minimal redundancy of these codes. We then derive the corresponding expressions for PB and CPB codes.
In Section~\ref{constructions}, we describe Knuth-like constructions for a variety of codes with various combinations of SB, CB, and PB properties.
Finally, the paper is concluded in Section~\ref{conc}.

\section{Preliminaries} \label{pre}

\subsection{Alphabets and Balancing}

\noindent
In Section~\ref{intro}, we introduced the alphabet $${\cal A}_q=\{-q+1,-q+3,-q+5,\ldots,q-3, q-1\},$$ where $q\ge 2$. We now formally define when a sequence ${\bf x}=(x_1,x_2,\ldots,x_n)\in({\cal A}_q)^n$ is balanced, for each of the considered perspectives.
\begin{itemize}
\item A sequence ${\bf x}$ of length $n=qm$, with $m \geq 1$, is \emph{symbol-balanced} (SB) if all $q$ symbols in ${\cal A}_q$ appear equally often in $\bf x$, i.e., $$|\{i: x_i=j\}|=m$$ for all $j\in{\cal A}_q$.
\item A sequence ${\bf x}$ of length $n$, with $n$ being a positive integer which is even if $q$ is even, is \emph{charge balanced} (CB) if the sum of all symbols in $\bf x$ is equal to $0$, i.e., $$\sum_{i=1}^n x_i=0.$$
\item A sequence ${\bf x}$ of length $n$, with $n$ being a positive integer which is even if $q$ is even, is \emph{polarity balanced} (PB) if the number of positive symbols in $\bf x$ equals the number of negative symbols, i.e., $$|\{i: x_i>0\}|=|\{i: x_i<0\}|.$$
\item A sequence ${\bf x}$ of length $n$, with $n$ being a positive integer which is even if $q$ is even, is \emph{charge and polarity balanced} (CPB) if it is both CB and PB.
\end{itemize}
Note that for lengths $n$ which do not comply with the specifications, there exist no sequences satisfying the desired property. Hence, throughout this paper, we will assume that $n$ is a multiple of $q$ for SB codes and that, in case $q$ is even, $n$ is even for CB, PB, and CPB codes.

When studying $q$-ary balanced codes, other alphabets than ${\cal A}_q$ have also been considered in the literature, a prominent example being $$\mathbb{Z}_q=\{0,1,\ldots, q-1\}.$$ Also balanced codes over the roots of unity alphabet $$\Phi_q=\{e^{2\pi i h/q}:h=0,1,\ldots,q-1\},$$ where $i=\sqrt{-1}$, have received quite some attention, e.g., \cite{BB2002}, \cite{MTAB2006}. The choice of the alphabet may influence the balancing notion. This is not the case for symbol balancing, which is clearly independent of symbol representation. The number of SB sequences of a certain length $n$ will be the same for any $q$-ary alphabet. The same conclusion is valid for polarity balancing, as long as we divide the alphabet symbols into two classes of equal size, with one neutral symbol in case $q$ is odd. However, the notion of charge balancing is coupled to the choice of the alphabet. First of all, it demands that an additive operation is defined on the alphabet symbols, which, by the way, does not have to be closed with respect to the alphabet, i.e., a sum of alphabet symbols may take values outside the alphabet. The naming {\em `charge'} and the choice to fix the sequence symbol sum $\sum_{i=1}^n x_i$ to zero, as in the CB definition above, have been inspired by practical PAM-like applications. However, in other cases it may be desirable to fix the sum to another value. Also, the maximum number of CB sequences of a certain length may depend on the choice of the alphabet: for an irregularly spaced alphabet other results could be obtained than for a regularly spaced alphabet like ${\cal A}_q$.

Throughout this paper, we will assume that the code alphabet is ${\cal A}_q$. Still, many derived results on maximum code sizes, minimum redundancies, etc., are also valid for other alphabets. Particularly, when the alphabet can be obtained by applying a bijective mapping of the format
\begin{equation}i\rightarrow ai+b\label{mapping}\end{equation}
on the symbols from ${\cal A}_q$, where $a\ne 0$ and $b$ are real numbers, then all results obtained for ${\cal A}_q$ also hold for the other alphabet (and vice versa), even the CB results. Note that $\mathbb{Z}_q$ is within this category (by choosing $a=-1/2$ and $b=(q-1)/2$). This implies that in $\mathbb{Z}_q$, the symbols smaller than $(q-1)/2$ should be called {\em `positive'} and the symbols larger than $(q-1)/2$ {\em `negative'}. Furthermore, the charge constraint should be replaced by $\sum_{i=1}^n x_i=n(q-1)/2$ in case the alphabet is $\mathbb{Z}_q$.

\subsection{Codes and Redundancy}

A \emph{code} of length $n$ is a set of sequences of length $n$. A code is said to be SB, CB, PB, or CPB if all codewords satisfy the respective properties. The sets of all SB, CB, PB, and CPB sequences of length $n$ over ${\cal A}_q$ are denoted by $C_{\rm SB}(n,q)$,  $C_{\rm CB}(n,q)$,  $C_{\rm PB}(n,q)$, and $C_{\rm CPB}(n,q)$, respectively, and their sizes by $M_{\rm SB}(n,q)$,  $M_{\rm CB}(n,q)$,  $M_{\rm PB}(n,q)$, and $M_{\rm CPB}(n,q)$, respectively. The \emph{redundancy} $r$ of a $q$-ary code of length $n$ and size $M$ is
\begin{equation} \label{red}
r=n-\log_qM.
 \end{equation}
The minimum redundancies of SB, CB, PB, and CPB codes of length $n$ over ${\cal A}_q$ are denoted by $r_{\rm SB}(n,q)$,  $r_{\rm CB}(n,q)$,  $r_{\rm PB}(n,q)$, and $r_{\rm CPB}(n,q)$, respectively.

\subsection{Stirling Approximation}

\noindent
In this paper, we will derive (asymptotic) expressions for the minimum redundancy. In the analysis we make frequent and implicit use of Stirling's approximation for factorials, stated here for convenience. For $n\ge 1$, it holds that
$$
n! = \sqrt{2 \pi n} \left(\frac{n}{e}\right)^n e^{\lambda_n}
$$
where $\frac{1}{12n+1} \leq \lambda_n \leq \frac{1}{12n}$. Hence,
\begin{equation} \label{stirlingO}
n! = \sqrt{2 \pi n} \left(\frac{n}{e}\right)^n \left(1+O\left(\frac{1}{n}\right)\right),
\end{equation}
and thus, for large values of $n$, we can use the approximation
\begin{equation}\label{stirling}
n! \approx \sqrt{2 \pi n} \left(\frac{n}{e}\right)^n.
\end{equation}

\subsection{Gaussian Approximation} \label{Gapp}

\noindent
Another tool which we will frequently use is the following Gaussian approximation technique.
We consider the symbols $x_i$ in a sequence $\bf x$ as $n$ independent random variables which are uniformly drawn from the alphabet ${\cal A}_q$.
We are interested in the distribution of the sum $\sum_{i=1}^n \phi(x_i)$, where $\phi$ is a function mapping symbols from ${\cal A}_q$ to real numbers, which has the property that the possible outcomes of the sum form a set of consecutive integer numbers.
Then, by the Central Limit Theorem, the probability that this sum takes the integer value $s$  is approximately
\begin{equation*} \frac {1} {\sigma\sqrt{2\pi}  }  e^{-\frac{1}{2}( \frac{s-\mu}{\sigma})^2},  \end{equation*}
with mean
\begin{equation} \mu=n  E[\phi(x)]=\frac{n}{q} \sum_{j=0}^{q-1} \phi(q-1-2j) \label{mean} \end{equation}
and variance
\begin{eqnarray}
\sigma^2&=&n  (E[(\phi(x))^2]-(E[\phi(x)])^2) \nonumber\\
&=&n\left(\left(\frac{1}{q} \sum_{j=0}^{q-1} (\phi(q-1-2j))^2\right)-\left(\frac{\mu}{n}\right)^2\right). \label{var}
\end{eqnarray}
Hence, the number of $q$-ary sequences of length $n$ with $\sum_{i=1}^n \phi(x_i)$ equal to $s$ is approximately
\begin{equation} \label{Gnum}
q^n  \frac{1}{\sigma\sqrt{2\pi}  }  e^{-\frac{1}{2}( \frac{s-\mu}{\sigma})^2}.
\end{equation}
Note that for fixed $n$ and $q$ this expression is maximum if $s$ is equal to $\mu$, which leads to a minimum redundancy of
$$\log_q \sigma + \frac{1}{2}\log_q 2\pi$$
when substituting (\ref{Gnum}) for $M$ in (\ref{red}).

\section{Minimum Redundancy of Balanced Codes}
\label{redundancy}

\noindent
In this section, we consider the cardinalities of $q$-ary SB, CB, PB, and CPB codes. From these cardinalities we derive asymptotic expressions for the minimum redundancies. The SB and CB results have been known for a long time but are reconsidered here for completeness. The PB and CPB results are new.

\subsection{Symbol-Balanced Sequences}

\noindent
For an SB code, all $q$ alphabet symbols must appear equally often in any codeword of length $n$. Hence, the problem of determining the number of such words boils down to a standard combinatorial problem. This number and the consequence with respect to minimum redundancy, as already discussed in \cite{MT2006}, are as follows.

\begin{theorem} \label{MSB}
For any $q$ and $n=mq$, it holds that
\begin{eqnarray*}M_{\rm SB}(n,q)&=&   {\frac{n!}{\left((n/q)!\right)^q}} \\
&\approx &  q^n\left(\frac{1}{2\pi n}\right)^{\frac{q-1}{2}}q^{\frac{q}{2}}.
\end{eqnarray*}
\end{theorem}
\beginofproof The equality follows from straightforward combinatorics and the approximation
from multiple uses of Stirling's formula (\ref{stirling}).
\endofproof

\begin{corollary}  \label{rSB}
For any $q$ and $n=mq$, it holds that
\begin{eqnarray*}
r_{\rm SB}(n,q) & = & n - \log_q M_{\rm SB}(n,q) \nonumber\\
  & \approx & \frac{q-1}{2}\log_q n + \frac{q-1}{2}\log_q 2\pi - \frac{q}{2}.
\end{eqnarray*}
\end{corollary}
\beginofproof The equality follows (by definition) from (\ref{red}) and the approximation from Theorem~\ref{MSB}.
\endofproof

By using (\ref{stirlingO}) rather than (\ref{stirling}), the more precise expressions
\begin{equation*}
M_{\rm SB}(n,q)=  q^n\left(\frac{1}{2\pi n}\right)^{\frac{q-1}{2}}q^{\frac{q}{2}}\left(1+O\left(\frac{1}{n}\right)\right)
\end{equation*}
and
\begin{equation*}
r_{\rm SB}(n,q)=  \frac{q-1}{2}\log_q n + \frac{q-1}{2}\log_q 2\pi - \frac{q}{2}+O\left(\frac{1}{n}\right)
\end{equation*}
are obtained. Hence, the approximation from Corollary~\ref{rSB} is exact if $n\rightarrow\infty$. This also holds for the approximate minimum redundancy expressions which will be presented in the subsequent subsections. In Subsection~\ref{subsecdisc}, we will illustrate the accuracy of the approximate expressions for finite values of $n$.

\subsection{Charge-Balanced Sequences}

\noindent
As observed by Capocelli {\em et al.} \cite{CGV1991} in their investigation of $q$-ary immutable codes, the number of words in a CB code of length $n$ was studied by Star \cite{S1975} in the context of his analysis of the number of restricted compositions of a positive integer. The final result is as stated in the next theorem, for which we provide a simple alternative proof.

\begin{theorem} \label{MCB}
For any $q$ and $n$ (which is even if $q$ is even), it holds that
$$M_{\rm CB}(n,q) \approx  q^n\sqrt{\frac{6}{\pi n(q^2-1)}}.$$
\end{theorem}
\beginofproof
We use the Gaussian approximation technique as discussed in Subsection~\ref{Gapp}.
Choosing the function $\phi$ to be
\begin{equation} \label{phiCB}
\phi(x)=\frac{x}{2},
\end{equation}
it follows that the number of sequences $\bf x$ over ${\cal A}_q$ of length $n$ with $\sum_{i=1}^n x_i=s$ is approximately equal to (\ref{Gnum}) with
mean
\begin{equation} \mu=\frac{n}{q} \sum_{j=0}^{q-1} \frac{q-1-2j}{2} =0 \label{meanCB} \end{equation}
(from (\ref{mean}) and (\ref{phiCB})) and variance
\begin{equation}\sigma^2=n\left(\frac{1}{q} \sum_{j=0}^{q-1} \left(\frac{q-1-2j}{2}\right)^2\right)=n\frac{q^2-1}{12}\label{varCB}\end{equation}
(from (\ref{var}), (\ref{phiCB}), and (\ref{meanCB})).
Note that CB sequences are characterized by the fact that $s=0$, and thus substitution of this value in (\ref{Gnum}), with $\mu=0$ and $\sigma^2=n(q^2-1)/12$, provides an approximation of $M_{\rm CB}(n,q)$. The result is as given in the theorem.
\endofproof

\begin{corollary}  \label{rCB}
For any $q$ and $n$ (which is even if $q$ is even), it holds that
\begin{eqnarray*}
r_{\rm CB}(n,q) &=& n -\log_q M_{\rm CB}(n,q) \\
  & \approx &  \frac{1}{2}\log_q n + \frac{1}{2}\log_q \frac{\pi(q^2-1)}{6}.
\end{eqnarray*}
\end{corollary}
\beginofproof The equality follows (by definition) from (\ref{red}) and the approximation from Theorem~\ref{MCB}.
\endofproof

\subsection{Polarity-Balanced Sequences}

\noindent
When calculating the number of $q$-ary PB sequences of length $n$, we distinguish between the cases $q$ is even and $q$ is odd, since in the latter case we should take into account the fact that the code alphabet contains the symbol `0' which is of indeterminate polarity. The results are presented in the next theorems, while expressions for the minimum redundancies of PB codes are given in the subsequent corollaries.

\begin{theorem} \label{MPBeven} For any even $q$ and even $n$, it holds that
\begin{eqnarray}M_{\rm PB}(n,q)&=& {n \choose {n/2}} \left(\frac{q}{2}\right)^n \label{eqMPBeven}\\
&\approx& q^n  \sqrt{\frac{2}{\pi n}}. \label{appMPBeven}
\end{eqnarray}
\end{theorem}
\beginofproof
The equality (\ref{eqMPBeven}) follows by observing that there are ${n \choose {n/2}}$ ways to create a balanced polarity pattern over $n$ positions and that for each such pattern we have $q/2$ symbol options for every positions. The approximation can be obtained by multiple uses of Stirling's formula (\ref{stirling}) or by applying the Gaussian approximation technique discussed in Subsection~\ref{Gapp}. Here, we opt for the latter, since intermediate results also turn out to be useful for the CPB case. Choosing the function $\phi$ to be
\begin{equation} \label{phiPBeven}
\phi(x) = \left \{ \begin{tabular}{ll}
$-\frac{1}{2}$,& if $x<0$,\\
$+\frac{1}{2}$, & if $x>0$,\end{tabular}\right.
\end{equation}
it follows that the number of $q$-ary sequences $\bf x$ of length $n$ with $\sum_{i=1}^n \phi(x_i)=s$ is approximately equal to (\ref{Gnum}) with
mean
\begin{equation} \mu=\frac{n}{q} \sum_{j=0}^{q-1} \phi(q-1-2j) =0\label{meanPBeven} \end{equation}
(from (\ref{mean}) and (\ref{phiPBeven})) and variance
\begin{equation}
\sigma^2=n\left(\frac{1}{q} \sum_{j=0}^{q-1} (\phi(q-1-2j))^2\right)=\frac{n}{4} \label{varPBeven}
\end{equation}
(from (\ref{var}), (\ref{phiPBeven}), and (\ref{meanPBeven})).
Note that PB sequences are characterized by the fact that $s=0$, and thus substitution of this value in (\ref{Gnum}), with $\mu=0$ and $\sigma^2=n/4$, gives (\ref{appMPBeven}).
\endofproof

\begin{corollary}  \label{rPBeven} For any even $q$ and even $n$, it holds that
\begin{eqnarray*}
r_{\rm PB}(n,q) &=& n -\log_q M_{\rm PB}(n,q) \\
  & \approx &  \frac{1}{2} \log_q n + \frac{1}{2} \log_q \frac{\pi}{2}.
\end{eqnarray*}
\end{corollary}
\beginofproof The equality follows (by definition) from (\ref{red}) and the approximation from Theorem~\ref{MPBeven}.
\endofproof

\begin{theorem} \label{MPBodd} For any $n$ and odd $q$, it holds that
\begin{eqnarray}M_{\rm PB}(n,q)&=& \sum_{j=0}^{\lfloor n/2 \rfloor} \frac{n!}{j!j!(n-2j)!}\left(\frac{q-1}{2}\right)^{2j} \label{eqMPBodd}\\
&\approx& q^n  \sqrt{\frac{q}{2\pi n(q-1)}}. \label{appMPBodd}
\end{eqnarray}
\end{theorem}
\beginofproof
The number of $q$-ary PB sequences of length $n$ with $j$ positive symbols, $j$ negative symbols, and thus $n-2j$ neutral symbols, is $\frac{n!}{j!j!(n-2j)!}\left(\frac{q-1}{2}\right)^{2j}$,  since there are $\frac{n!}{j!j!(n-2j)!}$ ways to create the postive/negative/neutral pattern over $n$ positions and for each such pattern we have $(q-1)/2$ symbol options for every non-neutral position. Summing over all possible values of $j$ shows (\ref{eqMPBodd}).

In order to obtain a simple expression for large values of $n$, we again use the Gaussian approximation technique introduced in Subsection~\ref{Gapp}. Proceeding as in the proof of Theorem~\ref{MPBeven}, while replacing the function $\phi$ by
\begin{equation} \label{phiPBodd}
\phi(x) = \left \{ \begin{tabular}{ll}
$-1$,& if $x<0$,\\
$0$,          & if $x=0$, \\
$+1$, & if $x>0$,\end{tabular}\right.
\end{equation}
giving mean
\begin{equation} \mu=\frac{n}{q} \sum_{j=0}^{q-1} \phi(q-1-2j) =0\label{meanPBodd} \end{equation}
(from (\ref{mean}) and (\ref{phiPBodd})) and variance
\begin{equation}
\sigma^2=n\left(\frac{1}{q} \sum_{j=0}^{q-1} (\phi(q-1-2j))^2\right)
=\frac{n(q-1)}{q} \label{varPBodd}
\end{equation}
(from (\ref{var}), (\ref{phiPBodd}) and (\ref{meanPBodd})),
we obtain (\ref{appMPBodd}).
\endofproof

\begin{corollary}  \label{rPBodd} For any $n$ and odd $q$, it holds that
\begin{eqnarray*}
r_{\rm PB}(n,q) &=& n -\log_q M_{\rm PB}(n,q) \\
  & \approx &  \frac{1}{2} \log_q n + \frac{1}{2} \log_q \frac{2\pi(q-1)}{q}.
\end{eqnarray*}
\end{corollary}
\beginofproof The equality follows (by definition) from (\ref{red}) and the approximation from Theorem~\ref{MPBodd}.
\endofproof

\subsection{Charge $\&$ Polarity-Balanced Sequences}

\noindent
Since each of the alphabets ${\cal A}_2=\{-1,+1\}$ and ${\cal A}_3=\{-2,0,+2\}$ has exactly one positive and one negative symbol, which have equal absolute value, it follows immediately from the definitions that the CB and PB constraints are completely equivalent for sequences over these alphabets. Therefore, for $q\le 3$, any CB sequence is also PB, and vice versa.

Hence, the minimum redundancy of a binary/bipolar CPB code of even length $n$ satisfies
\begin{eqnarray*}
r_{\rm CPB}(n,2) &=&r_{\rm CB}(n,2)=r_{\rm PB}(n,2)\\&\approx& \frac{1}{2} \log_2 n + \frac{1}{2} \log_2 \frac{\pi}{2},
\end{eqnarray*}
where the final expression follows from Corollary~\ref{rCB} or {\ref{rPBeven}. Furthermore, note that we have the same expression for $r_{\rm SB}(n,2)$; see Corollary~\ref{rSB}. This does not come as a surprise, as all balancing perspectives under consideration in the paper are equivalent in the binary/bipolar case.

For the minimum redundancy of a ternary CPB code of length $n$ we find
\begin{eqnarray*}
r_{\rm CPB}(n,3) &=&r_{\rm CB}(n,3)=r_{\rm PB}(n,3)\\&\approx& \frac{1}{2} \log_3 n + \frac{1}{2} \log_3 \frac{4\pi}{3},
\end{eqnarray*}
where the final expression follows from Corollary~\ref{rCB} or {\ref{rPBodd}. In this case, the corresponding expression for symbol balancing, provided by Corollary~\ref{rSB}, is
$$r_{\rm SB}(n,3)\approx \log_3 n + \log_3 2\pi-\frac{3}{2}, $$
which exceeds $r_{\rm CPB}(n,3)$ roughly by a factor of two.

As already argued in Section~\ref{intro}, the notions of CB and PB are not the same in case $q\ge 4$. First, we precisely determine, by combinatorial arguments, the number of CPB sequences of length $n$ in case $q=4$. Then, we derive approximate expressions for the number of CPB sequences for $q\ge 4$, from which we obtain the minimum redundancy.

We can count the number of CPB sequences over ${\cal A}_4$ of even length $n$ as follows. Polarity balancing requires that $n/2$ positions take values in $\{-3,-1\}$. If the number of such positions taking value $-3$ is $i$, then charge balancing requires that in the complementary set of $n/2$ positions taking values in $\{+1,+3\}$ there must be $i$ positions that take the value $+3$. Therefore, the size of  the intersection of the sets of CB and PB sequences is given by
\begin{eqnarray}
M_{\rm CPB}(n,4) &=& {n \choose {n/2}} \left(\sum_{i=0}^{n/2} {{n/2} \choose i} {{n/2} \choose i} \right) \nonumber\\
&=& {n \choose {n/2}} \left(\sum_{i=0}^{n/2} {{n/2} \choose i} {{n/2} \choose {(n/2)-i}}\right)\nonumber \\
&=& {n \choose {n/2}}{n \choose {n/2}}={n \choose {n/2}}^2. \label{MCPB4}
\end{eqnarray}

It seems to be cumbersome to extend the arguments used in the $q=4$ case to determine $M_{\rm CPB}(n,q)$ for larger values of $q$. However, the elegant Gaussian approximation method is still feasible, albeit that we need a joint distribution this time, since we have two constraints.
The results are presented in the next theorems and corollaries.

\begin{theorem} \label{MCPBeven} For any even $q\ge 4$ and even $n$, it holds that
$$M_{\rm CPB}(n,q)\approx q^n  \frac{1}{\pi n}\sqrt{\frac{48}{q^2-4}}.$$
\end{theorem}
\beginofproof
We consider the symbols $x_i$ in a sequence $\bf x$ as $n$ independent random variables which are uniformly drawn from the alphabet ${\cal A}_q$ with $q\ge 4$ even. We are interested in the joint distribution of the sums $S_1=\sum_{i=1}^n x_i/2$ and $S_2=\sum_{i=1}^n \phi(x_i)$, where $\phi$ is as defined in (\ref{phiPBeven}).
The probability that these sums take the integer values $s_1$ and $s_2$, respectively, is approximately
\begin{equation*} \frac {1} {2\pi\sigma_1\sigma_2\sqrt{1-\rho^2}}  e^{-\frac{1}{2(1-\rho^2)}f(s_1,s_2)},  \end{equation*}
where
$$f(s_1,s_2)=\sum_{i=1}^2\left(\frac{s_i-\mu_i}{\sigma_i}\right)^2
%+\left(\frac{s_2-\mu_2}{\sigma_2}\right)^2
-\frac{2\rho(s_1-\mu_1)(s_2-\mu_2)}{\sigma_1\sigma_2},$$
$$\mu_1=0 \mbox{ (from (\ref{meanCB}))}, $$
$$\sigma_1=\sqrt{\frac{n(q^2-1)}{12}} \mbox{ (from (\ref{varCB}))}, $$
$$\mu_2=0 \mbox{ (from (\ref{meanPBeven}))}, $$
$$\sigma_2=\sqrt{\frac{n}{4}} \mbox{ (from (\ref{varPBeven}))}, $$
and the correlation coefficient is
\begin{eqnarray*}
\rho&=&\frac{E[(S_1-\mu_1)(S_2-\mu_2)]}{\sigma_1\sigma_2} = \frac{E[S_1S_2]}{\sqrt{\frac{n(q^2-1)}{12}}\sqrt{\frac{n}{4}}}\\
&=& \frac{\frac{n}{2q}\sum_{i=0}^{\frac{q}{2}-1}(q-1-2i)}{n\sqrt{\frac{(q^2-1)}{48}}}
=\sqrt{\frac{3q^2}{4(q^2-1)}}.
\end{eqnarray*}
Hence, the number of $q$-ary sequences of length $n$ with $S_1=s_1$ and $S_2=s_2$ is approximately
\begin{equation} \label{Gnumjoint}
q^n  \frac {1} {2\pi\sigma_1\sigma_2\sqrt{1-\rho^2}}  e^{-\frac{1}{2(1-\rho^2)}f(s_1,s_2)}.
\end{equation}
Substitution of $s_1=0$ (the charge constraint), $s_2=0$ (the polarity constraint), and the two mean values, the two standard deviations, and the correlation coefficient, gives the stated result.
\endofproof

\noindent Note that this theorem gives
$$ M_{\rm CPB}(n,4)\approx 4^n  \frac{2}{\pi n},$$
a result which can also be obtained by applying the Stirling formula (\ref{stirling}) multiple times on (\ref{MCPB4}).

\begin{corollary}  \label{rCPBeven} For any even $q\ge 4$ and even $n$, it holds that
\begin{eqnarray*}
r_{\rm CPB}(n,q) &=& n -\log_q M_{\rm CPB}(n,q) \\
  & \approx  &  \log_q n + \log_q \left(\pi \sqrt{\frac{q^2-4}{48}}\right).
\end{eqnarray*}
\end{corollary}
\beginofproof The equality follows (by definition) from (\ref{red}) and the approximation from Theorem~\ref{MCPBeven}.
\endofproof

\begin{theorem} \label{MCPBodd} For any $n$ and odd $q\ge 5$, it holds that
$$M_{\rm CPB}(n,q)\approx q^n  \frac{1}{\pi n}\sqrt{\frac{12q^2}{(q^2-1)(q-1)(q-3)}}.$$
\end{theorem}
\beginofproof
We follow the same reasoning as in the proof of Theorem~\ref{MCPBeven},
though now using (\ref{phiPBodd}) instead of (\ref{phiPBeven}) for the $\phi$ function. Consequently, the standard deviation of $S_2$ changes to
$$\sigma_2=\sqrt{\frac{n(q-1)}{q}} \mbox{ (from (\ref{varPBodd}))}, $$
and the correlation coefficient to
\begin{eqnarray*}
\rho&=&\frac{E[(S_1-\mu_1)(S_2-\mu_2)]}{\sigma_1\sigma_2} = \frac{E[S_1S_2]}{\sqrt{\frac{n(q^2-1)}{12}}\sqrt{\frac{n(q-1)}{q}}}\\
&=& \frac{\frac{n}{2q}\sum_{i=0}^{\frac{q-3}{2}}(q-1-2i)}{n\sqrt{\frac{(q^2-1)(q-1)}{12q}}}
=\sqrt{\frac{3(q+1)}{4q}}.
\end{eqnarray*}
The final result follows by substituting all the parameters in (\ref{Gnumjoint}).
\endofproof

\begin{corollary}  \label{rCPBodd} For any $n$ and odd $q\ge 5$, it holds that
\begin{eqnarray*}
r_{\rm CPB}(n,q) &=& n -\log_q M_{\rm CPB}(n,q) \\
  & \approx  &  \log_q n + \\&& \log_q \left(\pi \sqrt{\frac{(q^2-1)(q-1)(q-3)}{12q^2}}\right).
\end{eqnarray*}
\end{corollary}
\beginofproof The equality follows (by definition) from (\ref{red}) and the approximation from Theorem~\ref{MCPBodd}.
\endofproof

\subsection{Discussion} \label{subsecdisc}

\noindent
In this subsection, we discuss the results on the minimum redundancy of balanced codes as obtained in this section.
As stated before, the minimum redundancy expressions as presented in the corollaries are approximations which are exact if $n\rightarrow\infty$. For finite values of $n$, the accuracy of these expressions depends on the convergence rates of the underlying Stirling/Gaussian  approximations. Here, we provide an illustration by showing some numerical values for $r_{\rm CPB}(n,4)$, i.e., the minimum redundancy of a CPB code of length $n$ over ${\cal A}_4$. From (\ref{red}) and (\ref{MCPB4}) we obtain the exact expression
\begin{equation} \label{cpb4exact}
r_{\rm CPB}(n,4) = n -2\log_4 {n \choose {n/2}},
\end{equation}
while Corollary~\ref{rCPBeven} gives the approximate expression
\begin{equation}  \label{cpb4app}
r_{\rm CPB}(n,4)\approx \log_4 (n\pi/2).
\end{equation}
The comparison of these two expressions as given in Table~\ref{tabcomp4} shows that the approximation is quite accurate, even for small values of $n$.
\begin{table}
\caption{Numerical Values for $r_{\rm CPB}(n,4)$}
\label{tabcomp4}
$$\begin{array}{|c|c|c|} \hline
n & \mbox{Exact, Eq. $(\ref{cpb4exact})$} & \mbox{Approximation, Eq. $(\ref{cpb4app})$} \\ \hline
10   & 2.0227 & 1.9867 \\
20   & 2.5047 & 2.4867 \\
40   & 2.9957 & 2.9867 \\
60   & 3.2852 & 3.2792 \\
80   & 3.4912 & 3.4867 \\
100  & 3.6513 & 3.6477 \\
200  & 4.1495 & 4.1477 \\
400  & 4.6486 & 4.6477 \\
600  & 4.9408 & 4.9402 \\
800  & 5.1481 & 5.1477 \\
1000 & 5.3090 & 5.3086 \\ \hline
\end{array}$$
\end{table}

Note that all minimum redundancy expressions are of the form
$$g(q) \log_q n + h(q),$$
where $g$ and $h$ are functions such that the output values may depend on the alphabet size $q$ but not on the block length $n$. For comparison purposes, we introduce the {\em asymptotic normalized redundancy} (ANR) as the redundancy divided by $\log_q n$ in the limit of large values of $n$. Note that this ANR is equal to $g(q)$.
For example, it follows from Corollary~\ref{rSB} that
$$ g_{\rm SB}(q)=\frac{q-1}{2}.$$
The complete overview of these ANRs is provided in Table~\ref{normred}.
\begin{table}
\caption{Asymptotic normalized redundancies}
\label{normred}
$$\begin{array}{|c|c|c|c|c|} \hline
  & {\rm SB} & {\rm CB}& {\rm PB}& {\rm CPB} \\ \hline &&&& \\
q=2&\frac{1}{2}&\frac{1}{2}&\frac{1}{2}&\frac{1}{2}\\ &&&& \\\hline &&&& \\
q=3&1&\frac{1}{2}&\frac{1}{2}&\frac{1}{2}\\ &&&& \\\hline &&&& \\
q\ge 4&\frac{q-1}{2}&\frac{1}{2}&\frac{1}{2}&1\\ &&&& \\\hline
\end{array}$$
\end{table}
From this table, we conclude that the CB and PB properties are equally expensive in terms of ANR, while the SB property is $q-1$ times as expensive. The combined CB and PB property (CPB) is as expensive as either of the individual properties, i.e., the other comes for free, if $q\le 3$, while it costs the sum of the individual contributions if $q\ge 4$.

\section{Constructions of Balanced Codes}
\label{constructions}

\noindent
In the previous section we have determined expressions for the number $M(n,q)$ of $q$-ary sequences of length $n$ satisfying certain balancing constraints. From these expressions we calculated the minimum required code redundancy to achieve the constraints. However, the lists of balanced words come with little structure. Applying table look-up is only feasible for small codes, but for practical implementation of larger codes, we need simple encoding and decoding algorithms. Knuth presented such an algorithm for the case $q=2$, i.e., for binary/bipolar balanced codes \cite{K1986}. Here, we will propose extensions to non-binary codes from various balancing perspectives.

All proposed methods take an approach similar to the original Knuth construction. We make simple and reversible modifications to a $q$-ary information sequence $\bf u$ of length $k$ to obtain a $q$-ary balanced sequence $\bf x$ of the same length. Next, we create a $q$-ary balanced prefix $\bf p$ of length $p$, which uniquely identifies the modifications. The $q$-ary balanced codeword ${\bf c}=({\bf p},{\bf x})$ of length $n=p+k$ is then transmitted or stored. The receiver retrieves the modifications from the prefix and applies these in reverse on $\bf x$ to obtain the original $\bf u$.

The constructions are nice and simple, but not optimal with respect to redundancy. Note that all codewords consist of two parts which are both balanced, and thus words which are balanced overall, but not within these parts, are excluded. Hence, simplicity comes at a price of increased redundancy. In order to still keep the redundancy as small as possible within the construction framework, we should minimize the prefix length $p$. Since the prefix is much shorter than the information sequence, we will assume that encoding and decoding of the prefix can be done by table look-up or another minimum redundancy achieving method. Let the number of different prefixes required to uniquely identify the modifications be denoted by $P$. Ignoring balancing, the number of $q$-ary symbols needed to represent the prefix is thus
\begin{equation} \label{pp}
p'=\log_q P,
\end{equation}
which we will call the \emph{unbalanced redundancy}. The actual prefix length will be (a little bit) larger, since the prefix needs to be balanced as well. It should be chosen as the smallest integer $p$ such that
\begin{equation} \label{pe}
M(p,q)\ge P.
\end{equation}
The analysis from the previous section shows that, for fixed $q$, the extra redundancy to make the prefix balanced is in the order of $\log p'$, i.e.,
$$p=p'+O(\log p').$$
Hence, for rough evaluation purposes, the unbalanced redundancy $p'$, which is easily determined by (\ref{pp}), may serve as a satisfactory approximation of the actual redundancy $p$, which requires the more cumbersome computation from (\ref{pe}).

%In the description of the algorithms, we will frequently use the so-called running digital sum ${\rm RDS}_j({\bf y})$ of a $q$-ary sequence ${\bf %y}=(y_1,y_2,...,y_k)$, which is defined by
%$${\rm RDS}_j({\bf y})=\sum_{i=1}^j y_i$$
%for $j=0,1,\ldots,k$, where the summation is over the real numbers. Note that ${\rm RDS}_0({\bf y})=0$ and that ${\rm RDS}_k({\bf y})$ is the sum of %all symbols in $\bf y$.
All constructions will be presented for the code alphabet ${\cal A}_q$, but equivalents for other alphabets, e.g., $\mathbb{Z}_q$, can be established using the mapping (\ref{mapping}).
Before starting the descriptions of the constructions, we introduce some more notation. The real sum of all symbols in a $q$-ary sequence $\bf y$ is denoted by ${\rm Sum}({\bf y})$, i.e.,
$${\rm Sum}({\bf y})=\sum_i y_i.$$
Further, let $S_j({\bf y})$ denote the number of appearances of the alphabet symbol $j$ in ${\bf y}$, i.e.,
$$S_j({\bf y})=|\{i:y_i=j\}$$
for any alphabet symbol $j$.
Finally, as a short-hand notation, we denote a run of $b$ symbols $a$ by $a^b$, e.g., $3^2 1^3 (-1)^1 3^2$ denotes the sequence $(3,3,1,1,1,-1,3,3)$.

\subsection{Knuth's Construction} \label{subsecKnuth}

\noindent
We start by stating Knuth's original construction for bipolar codes \cite{K1986}, as a reference. For any information sequence $\bf u$ of even length $k$ and any $j\in\{0,1,\ldots,k\}$, let ${\bf u}'_j$ denote the sequence $\bf u$ with the first $j$ symbols multiplied by $-1$. A {\em balancing index} is a number $z$ for which ${\bf u}'_z$ is balanced.

\noindent{\bf Knuth Encoding Procedure}
\begin{enumerate}
\item Determine a balancing index $z\in\{0,1,\ldots,k-1\}$ for the information sequence $\bf u$.
\item Multiply the first $z$ symbols of $\bf u$ by $-1$ to obtain the balanced sequence $\bf x$.
\item Map $z$ to a unique balanced prefix $\bf p$.
\end{enumerate}
Then transmit or store the balanced codeword ${\bf c}=({\bf p},{\bf x})$.

\noindent{\bf Knuth Decoding Procedure}
\begin{enumerate}
\item Retrieve the balancing index $z$ from $\bf p$.
\item Multiply the first $z$ symbols of $\bf x$ by $-1$ to retrieve $\bf u$.
\end{enumerate}
\beginofproof
It is easy to see that the operation in the encoding procedure is properly reversed in the decoding procedure. Hence, we only need to show that for every sequence $\bf u$ of length $k$ there
exists at least one $z\in\{0,1,\ldots,k-1\}$ such that ${\bf u}'_z$ is balanced, i.e., ${\rm Sum}({{\bf u}'_z})=0$. This immediately follows from combining the following observations.
\begin{enumerate}
\item ${\rm Sum}({\bf u}'_0)$ is even.
\item ${\rm Sum}({\bf u}'_{j})={\rm Sum}({\bf u}'_{j-1})\pm 2$ for all $j\in\{1,2,\ldots,k\}$.
\item ${\rm Sum}({\bf u}'_k)=-{\rm Sum}({\bf u}'_0)$.
\end{enumerate}
\endofproof

Since there are $k$ possible values for $z$, the redundancy, i.e., the length $p$ of the prefix, is a little bit more than $p'=\log_2 k$.

\begin{example} \label{exKnuth}
For the bipolar sequence $${\bf u}=(+1,-1,+1,+1,+1,+1)$$ of length 6, encoding goes as follows.
\begin{enumerate}
\item Find the balancing index to be $z=4$.
\item Invert the first $4$ positions of $\bf u$, i.e., $${\bf x}=(-1,+1,-1,-1,+1,+1).$$
\item Uniquely map the balancing index $4$ to one of the six balanced sequences of length four, e.g., $${\bf p}=(+1,-1,-1,+1).$$
\end{enumerate}
Then the balanced transmitted/stored sequence is $${\bf c}=({\bf p},{\bf x})=(+1,-1,-1,+1,-1,+1,-1,-1,+1,+1).$$
\end{example}

\subsection{Polarity-Balanced Code Construction} \label{subsecKnuthPB}

\noindent
Knuth's original method for generating balanced binary sequences can be adapted to generate $q$-ary PB sequences. This is rather straightforward, although there is a snag if $q$ is odd. In this case, the number of zero-valued symbols in $\bf u$ may be of different parity than the length $k$, which results in an odd number of non-zero (either positive or negative) symbols. Since the value zero is (polarity-)neutral, i.e., neither positive nor negative, inversion of any number of symbols in $\bf u$ will not lead to a PB sequence in such a situation. We will solve this by introducing an offset in case $q$ is odd.
We propose the following algorithm for sequences over ${\cal A}_q$, where $\oplus_{2q}$ denotes the addition over the integer numbers, with a reduction modulo $2q$ such that the final outcome is in ${\cal A}_q$.

\noindent{\bf PB Encoding Procedure}
\begin{enumerate}
\item If $q$ is odd, then determine a symbol $a$ in ${\cal A}_q$ such that $S_a({\bf u})$ has the same parity as the length $k$ of $\bf u$, i.e., $S_a({\bf u})$ and $k$ are either both even or both odd.
\item If $q$ is odd, then compute ${\bf u}'={\bf u}\oplus_{2q} (-{\bf a})$, where ${\bf a}=(a,a,\ldots,a)$ is of length $k$. If $q$ is even, then ${\bf u}'={\bf u}$.
\item Determine a polarity balancing index $z\in\{0,1,\ldots,k-1\}$ for ${\bf u}'$.
\item Multiply the first $z$ positions of ${\bf u}'$ by $-1$ to obtain the PB sequence $\bf x$.
\item Map $z$ (if $q$ is even) or $(a,z)$ (if $q$ is odd) to a unique PB prefix $\bf p$.
\end{enumerate}
Then transmit or store the balanced codeword ${\bf c}=({\bf p},{\bf x})$.

\noindent{\bf PB Decoding Procedure}
\begin{enumerate}
\item Retrieve the balancing index $z$ from $\bf p$.
\item Multiply the first $z$ positions of $\bf x$ by $-1$ to retrieve $\bf u$ (if $q$ is even) or ${\bf u}'$ (if $q$ is odd).
\item If $q$ is odd, then retrieve $a$ from the prefix $\bf p$ and compute ${\bf u}={\bf u}'\oplus_{2q} {\bf a}$.
\end{enumerate}
\beginofproof
It is easy to see that the operations in the encoding procedure are properly reversed in the decoding procedure. Hence, we only need to show the existence of (i) a suitable offset $a$ (in case $q$ odd) and (ii) a suitable polarity balancing index $z$.

(i) The existence of $a$ can be demonstrated by supposing it does not exist and then deriving a contradiction. If $q$ and $k$ are odd, then $S_j({\bf u})$ is odd for at least one symbol $j\in{\cal A}_q$, since all of them being even would imply that $k=\sum_i S_i({\bf u})$ is even. If $q$ is odd and $k$ is even, then $S_j({\bf u})$ is even for at least one $j\in{\cal A}_q$, since all of them being odd would imply that $k=\sum_i S_i({\bf u})$, a summation of an odd number of odd terms, is odd.

(ii) The existence of $z$ follows by a similar argument as for the Knuth algorithm.
Let ${\bf u}'_j$ denote the sequence ${\bf u}'$ with the first $j$ symbols multiplied by $-1$ and let $\phi$ be defined as in (\ref{phiPBodd}). For a PB balancing index $z$, it must hold that ${\rm Sum}(\phi({\bf u}'_j))=0$.
The existence of a PB balancing index follows by combining the following observations.
\begin{enumerate}
\item ${\rm Sum}(\phi({\bf u}'_0))$ is even, since the number of non-zero symbols in ${\bf u}'$ is even.
\item ${\rm Sum}(\phi({\bf u}'_j))={\rm Sum}(\phi({\bf u}'_{j-1}))+c$ for all $j\in\{1,2,\ldots,k\}$, where $c\in\{-2,0,+2\}$.
\item ${\rm Sum}(\phi({\bf u}'_k))=-{\rm Sum}(\phi({\bf u}'_0))$.
\end{enumerate}
\endofproof

Since there are $k$ possible values for $z$ and $q$ possible values for $a$, we have $p'=\log_q k$ if $q$ is even and $p'=1+\log_q k$ if $q$ is odd.

\begin{example} \label{exPB}
Let $q=5$. For the sequence $${\bf u}=(+4,+4,-2,0,0,0,0)\in({\cal A}_5)^7,$$ encoding goes as follows.
\begin{enumerate}
\item Since $q=5$ and $k=7$ are odd, identify `$-2$' as the symbol $a$ with an odd number of appearances in $\bf u$.
\item Subtract (modulo 10) the value -2 from every symbol in $\bf u$, resulting in $${\bf u}'=(-4,-4,0,+2,+2,+2,+2).$$
\item Find the PB index $z$ to be $6$.
\item Multiply the first $6$ positions of ${\bf u}'$ by $-1$ to obtain $${\bf x}=(+4,+4,0,-2,-2,-2,+2).$$
\item Uniquely map $(a,z)=(-2,6)$ to one of the PB sequences of length $4$, e.g., $${\bf p}=(+2,0,0,-4).$$
\end{enumerate}
Then the balanced transmitted/stored sequence is $${\bf c}=(+2,0,0,-4,+4,+4,0,-2,-2,-2,+2).$$
\end{example}

\subsection{Charge-Balanced Code Construction} \label{subsecKnuthCB}

\noindent
In \cite{SW2009}, Swart and Weber presented a Knuth-like construction for $q$-ary CB codes over the alphabet $\mathbb{Z}_q$. We include it here, in a version for the alphabet ${\cal A}_q$, to make this paper self-contained. Furthermore, we need it in the subsequent subsection as a component for CPB code construction. The key ingredient of the CB method is a set of $qk$ balancing sequences ${\bf b}_i$, $i=0,1,\ldots,qk-1$, each consisting of $g$ symbols $j+2$ followed by $k-g$ symbols $j$, i.e.,
$${\bf b}_i=(j+2)^g j^{k-g},$$
where $j=2\lfloor i/k\rfloor$ and $g=i-k\lfloor i/k\rfloor$. Again, $\oplus_{2q}$ denotes the addition over the integer numbers, with a reduction modulo $2q$ such that the final outcome is in ${\cal A}_q$. A charge balancing index is a number $z$ such that ${\rm Sum}({\bf u}\oplus_{2q}{\bf b}_z)=0$.
The algorithm is described as follows.

\noindent{\bf CB Encoding Procedure}
\begin{enumerate}
\item Determine a CB index $z\in\{0,1,\ldots,qk-1\}$ for the information sequence $\bf u$.
\item Compute the CB sequence ${\bf x}={\bf u}\oplus_{2q}{\bf b}_z$.
\item Map $z$ to a unique CB prefix $\bf p$.
\end{enumerate}

\noindent Then transmit or store the balanced codeword ${\bf c}=({\bf p},{\bf x})$.

\noindent{\bf CB Decoding Procedure}
\begin{enumerate}
\item Retrieve the balancing index $z$ from $\bf p$.
\item Compute ${\bf u}={\bf x}\oplus_{2q}(-{\bf b}_z)$.
\end{enumerate}
\beginofproof
It is easy to see that the operation in the encoding procedure is properly reversed in the decoding procedure. Hence, we only need to show the existence of a CB index for  any information sequence $\bf u$ of length $k$. Define ${\bf b}_{qk}={\bf b}_0$, and consider the series $${\rm Sum}({\bf u}\oplus_{2q}{\bf b}_0), {\rm Sum}({\bf u}\oplus_{2q}{\bf b}_{1}), \ldots, {\rm Sum}({\bf u}\oplus_{2q}{\bf b}_{qk}).$$ We make the following observations.
\begin{enumerate}
\item The series starts and ends with the same even value.
\item For all $i\in\{0,1,\ldots,qk-1\}$, it holds that $${\rm Sum}({\bf u}\oplus_{2q}{\bf b}_{i+1})={\rm Sum}({\bf u}\oplus_{2q}{\bf b}_{i})+c,$$ where $c$ is either $+2$ or $-2q+2$.
\item It holds that
\begin{eqnarray*}\sum_{j=0}^{q-1}{\rm Sum}({\bf u}\oplus_{2q}{\bf b}_{jk})&=&\sum_{l=1}^{k}\sum_{j=0}^{q-1} (u_l\oplus_{2q}2j)\\
&=&k\sum_{j=0}^{q-1}(-q+1+2j)=0,
\end{eqnarray*}
 where the first equality follows from the fact that the sequence ${\bf b}_{jk}$ consists of $k$ symbols $2j$, and the second equality from the consequence that every position $l$ takes every symbol value from the alphabet ${\cal A}_q$ exactly once in the summation. Hence, the average value of all ${\rm Sum}({\bf u}\oplus_{2q}{\bf b}_{jk})$, with $j=0,1,\ldots,q-1$, is $0$.
\end{enumerate}
By combining these three observations, we can conclude that there exists at least one $z$ in $\{0,1,\ldots,qk-1\}$ such that ${\rm Sum}({\bf u}\oplus_{2q}{\bf b}_{z})=0$.
\endofproof

Since there are $qk$ possible values for $z$, the unbalanced redundancy is $p'=1+\log_q k$. Note that by setting $q=2$, we do not exactly get the original Knuth method as described in Subsection~\ref{subsecKnuth}, where $p'$ is one bit less. The reason is that for the binary case, it can be shown (as done by Knuth and in Subsection~\ref{subsecKnuth}) that there is always a suitable balancing index in a set of $k$ candidates (rather than $2k$). For further details, see \cite{SW2009}. Pelusi {\em et al.} \cite{PTB2010} presented a slightly improved $q$-ary CB coding scheme, using $(q-1)k+q \mod 2$ rather than $qk$ balancing functions, with the same asymptotic redundancy though.

\begin{example} \label{exCB}
We use the same information sequence as in Example~\ref{exPB}, i.e., $${\bf u}=(+4,+4,-2,0,0,0,0)\in({\cal A}_5)^7.$$ Encoding into a CB sequence goes as follows.
\begin{enumerate}
\item Find a suitable CB index $z$ to be $32$.
\item Compute the CB sequence
\begin{eqnarray*}{\bf x}&=& {\bf u}\oplus_{10}({\bf b}_{32})\\
&=& (+4,+4,-2,0,0,0,0) \oplus_{10} \\
&& (10,10,10,10,8,8,8)\\
&=& (+4,+4,-2,0,-2,-2,-2) \\
\end{eqnarray*}
\item Uniquely map the CB index 32 to one of the CB sequences of length $4$, e.g., $${\bf p}=(+4,0,-2,-2).$$
\end{enumerate}
Then the balanced transmitted/stored sequence is $${\bf c}=(+4,0,-2,-2,+4,+4,-2,0,-2,-2,-2).$$
Note that the sequence $\bf x$ generated this way is not PB. Rather than $z=32$, we could also have chosen $z=7$, but also then the resulting CB sequence
$${\bf x}=(-4,-4,0,+2,+2,+2,+2)$$
is not PB.
\end{example}

\subsection{Charge $\&$ Polarity-Balanced Code Construction} \label{subsecKnuthCPB}

\noindent
If $q\le 3$, then any code which is PB is also CB and vice versa. Hence, either of the coding strategies described in the previous two subsections provides CPB codes. However, for $q\ge 4$, the CB and PB properties are no longer equivalent, and a dedicated construction method is needed. Such a method
will be proposed in this subsection, where we will assume throughout that $q\ge 4$ and that $k$ is even if $q$ even.

For constructing codes having both the charge and polarity balancing properties, we can still base our constructions on the methods described in the previous two subsections. However, the straightforward strategy of first applying one method and then the other could fail, since the property obtained in the first round might be destroyed in the second. Therefore, a more sophisticated strategy should be developed.

In the proposed method, we first transform the information sequence $\bf u$ into a PB sequence as described in Subsection~\ref{subsecKnuthPB}. In this PB sequence, which we denote by $\bf y$, we focus on the subsequences ${\bf y}^+$, which consists of all positive symbols in $\bf y$,  and ${\bf y}^-$, which consists of all negative symbols. Both subsequences have the same length (due to the established PB property) which we denote by $k'$. Note that
$${\rm Sum}({\bf y}^-)\le 0\le {\rm Sum}({\bf y}^+).$$
We are going to make modifications to $\bf y$, affecting only ${\bf y}^+$ and ${\bf y}^-$, such that the resulting sequence $\bf x$ satisfies
\begin{equation} \label{SSS}
{\rm Sum}({\bf x}^+)+{\rm Sum}({\bf x}^-)=0,
\end{equation}
which implies that $\bf x$ is CPB.

The modifications are done in such a way that the polarity of all involved symbols will not change. Hence, like $\bf y$, the sequence $\bf x$ is PB.
The first step of the modification process consists of a possible `mirror' operation on the symbols in ${\bf y}^+$ (with respect to the value $\lceil q/2\rceil$).
Define
\begin{equation} \label{defxi}
\xi = \left \{ \begin{tabular}{ll}
$1$,& if ${\rm Sum}({\bf y}^+)<k'\lceil q/2\rceil<-{\rm Sum}({\bf y}^-)$ \\
&or $-{\rm Sum}({\bf y}^-)<k'\lceil q/2\rceil<{\rm Sum}({\bf y}^+)$,\\
$0$, & otherwise.\end{tabular}\right.
\end{equation}
If $\xi=1$, then all symbols $y_i$ in ${\bf y}^+$ are replaced by $2\lceil q/2\rceil-y_i$; else they are left untouched. Note that for the sequence $\bf z$ obtained from $\bf y$ by this operation, it holds that ${\rm Sum}({\bf z}^+)$ and $-{\rm Sum}({\bf z}^-)$ are both at least equal to $k'\lceil q/2\rceil$ or both at most equal to this value.
Define
\begin{equation} \label{defnu}
\nu = \left\{ \begin{tabular}{ll}
$+$,& if ${\rm Sum}({\bf z}^+)\ge-{\rm Sum}({\bf z}^-)\ge k'\lceil q/2\rceil$ \\
&or ${\rm Sum}({\bf z}^+)\le-{\rm Sum}({\bf z}^-)\le k'\lceil q/2\rceil$,\\
$-$, & otherwise.\end{tabular}\right.
\end{equation}

In the second (and last) step of the modification process, we change either the positive or the negative symbols in $\bf z$, in a manner similar to that used in Subsection~\ref{subsecKnuthCB}. Consider $\lfloor q/2\rfloor k'$ balancing sequences
$${\bf b}_i=(j+2)^g j^{k'-g},$$
$i=0,1,\ldots,\lfloor q/2\rfloor k'-1$, where $j=2\lfloor i/k'\rfloor$ and $g=i-k'\lfloor i/k'\rfloor$. Throughout the rest of this subsection, let $\oplus$ denote the addition over the integer numbers, with a reduction modulo $2\lfloor q/2\rfloor$ such that the final outcome is in ${\cal A}_q^+=\{j\in{\cal A}_q:j>0\}$ if $\nu=+$ and in ${\cal A}_q^-=\{j\in{\cal A}_q:j<0\}$ if $\nu=-$. We replace ${\bf z}^\nu$ by ${\bf z}^\nu\oplus{\bf b}_w$, where $w$ is chosen such that
\begin{equation}\label{zz}
{\rm Sum}({\bf z}^\nu\oplus{\bf b}_w)=
-{\rm Sum}({\bf z}^{\bar\nu}),
\end{equation}
where $\bar\nu$ denotes the inverse of $\nu$.
In conclusion, the resulting sequence $\bf x$ satisfies (\ref{SSS}).

In summary, we have the following algorithm in case $q\ge 4$.

\noindent{\bf CPB Encoding Procedure}
\begin{enumerate}
\item Apply the encoding procedure from Subsection~\ref{subsecKnuthPB} to change the information sequence $\bf u$ into a PB sequence $\bf y$, using appropriate offset $a$ (if $q$ is odd) and PB index $z$.
\item Compute $\xi$ by (\ref{defxi}).
\item If $\xi=1$, then replace all symbols $y_i$ in ${\bf y}^+$ by $2\lceil q/2\rceil-y_i$; else leave them untouched. Call the resulting sequence $\bf z$.
\item Compute $\nu$ by (\ref{defnu}).
\item Determine an index $w$ such that (\ref{zz}) is satisfied.
\item Replace in $\bf z$ the subsequence ${\bf z}^\nu$ by ${\bf z}^\nu\oplus{\bf b}_w$, to obtain the CPB sequence $\bf x$, .
\item Map $(z,\xi,\nu,w)$ (if $q$ even) or $(a,z,\xi,\nu,w)$ (if $q$ odd) to a unique CPB prefix $\bf p$.
\end{enumerate}

\noindent Then transmit or store the balanced codeword ${\bf c}=({\bf p},{\bf x})$.

\noindent{\bf CPB Decoding Procedure}
\begin{enumerate}
\item Retrieve $a$ (if $q$ is odd), $z$, $\xi$, $\nu$, and $w$ from the prefix $\bf p$.
\item Replace ${\bf x}^\nu$ by ${\bf x}^\nu\oplus(-{\bf b}_w)$ in $\bf x$ to obtain $\bf z$.
\item If $\xi=1$, then replace all symbols $z_i$ in ${\bf z}^+$ by $2\lceil q/2\rceil-z_i$; else leave them untouched. Call the resulting sequence $\bf y$.
\item Apply the decoding procedure from Subsection~\ref{subsecKnuthPB} to retrieve $\bf u$ from $\bf y$, using $a$ (if $q$ is odd) and $z$.
\end{enumerate}
\beginofproof
It is easy to see that the operations in the encoding procedure are properly reversed in the decoding procedure. Further, the validity of the PB part was already demonstrated in Subsection~\ref{subsecKnuthPB}. Hence, the only thing left to prove is that there always exists a suitable index $w$.
To this end, define ${\bf b}_{\lfloor q/2\rfloor k'}={\bf b}_0$ and consider the series $${\rm Sum}({\bf z}^\nu\oplus{\bf b}_0), {\rm Sum}({\bf z}^\nu\oplus{\bf b}_{1}), \ldots, {\rm Sum}({\bf z}^\nu\oplus{\bf b}_{\lfloor q/2\rfloor k'}).$$ We make the following observations.
\begin{enumerate}
\item The series starts and ends with the same even value.
\item For all $i\in\{0,1,\ldots,\lfloor q/2\rfloor k'-1\}$, it holds that $${\rm Sum}({\bf z}^\nu\oplus{\bf b}_{i+1})={\rm Sum}({\bf z}^\nu\oplus{\bf b}_{i})+c,$$ where $c$ is either $+2$ or $-2\lfloor q/2\rfloor+2$.
\item It holds that
\begin{eqnarray*}
\left|\sum_{j=0}^{\lfloor q/2\rfloor-1}{\rm Sum}({\bf z}^\nu\oplus{\bf b}_{jk'})\right|
&=&k'\sum_{j=0}^{\lfloor q/2\rfloor-1}(q-1-2j)\\
&=&\lfloor q/2\rfloor k' \lceil q/2\rceil.
\end{eqnarray*}
Hence, the average value of all ${\rm Sum}({\bf z}^\nu\oplus{\bf b}_{jk'})$, with $j=0,1,\ldots,\lfloor q/2\rfloor-1$, is $k' \lceil q/2\rceil$.
\end{enumerate}
By combining these three observations and (\ref{defnu}), we can conclude that there exists at least one $w$ in $\{0,1,\ldots,\lfloor q/2\rfloor k'-1\}$ such that (\ref{zz}) is satisfied.
\endofproof

Since there are $q$ possible values for $a$, $k$ for $z$, $2$ for $\xi$, $2$ for $\nu$, and $\lfloor q/2\rfloor k'\le \lfloor q/2\rfloor\lfloor k/2\rfloor$ for $w$, it is sufficient to choose the prefix length such that
$$P= 4k\lfloor q/2\rfloor\lfloor k/2\rfloor=qk^2$$
CPB sequences can be accommodated if $q$ is even, and
$$P= 4qk\lfloor q/2\rfloor\lfloor k/2\rfloor=2q(q-1)k\lfloor k/2\rfloor$$
if $q$ is odd. Hence, the unbalanced redundancy is $$p'=\log_q P=1+2\log_q k$$ if $q$ is even, and very close to that number if $q$ is odd.

\begin{example} \label{exCPB}
We use the same information sequence as in Examples~\ref{exPB} and \ref{exCB}, i.e., $${\bf u}=(+4,+4,-2,0,0,0,0)\in({\cal A}_5)^7.$$ Encoding into a CPB sequence goes as follows.
\begin{enumerate}
\item From Example~\ref{exPB}, the PB sequence $${\bf y}=(+4,+4,0,-2,-2,-2,+2)$$ is obtained.
\item Find $\xi=1$, since $$-{\rm Sum}({\bf y}^-)=6<9<10={\rm Sum}({\bf y}^+).$$
\item Mirror the positive values in $\bf y$ with respect to $+3$ to obtain
$${\bf z}=(+2,+2,0,-2,-2,-2,+4).$$
\item Find $\nu=-$, since $$-{\rm Sum}({\bf z}^-)=6<8={\rm Sum}({\bf z}^+)\le 9.$$
\item Determine $w=1$ as a suitable balancing index.
\item Add (modulo 4, with the resulting symbols in the set $\{-4,-2\}$) the sequence
${\bf b}_1=(2,0,0)$ to ${\bf z}^-$, i.e., compute
\begin{eqnarray*}{\bf x}&=&
(+2,+2,0,-2,-2,-2,+4) \\
&& \oplus(0,0,0,2,0,0,0)\\
&=& (+2,+2,0,-4,-2,-2,+4)
\end{eqnarray*}
\item Uniquely map $(a,z,\xi,\nu,w)=(-2,6,1,-,1)$ to one of the CPB sequences of length $6$, e.g., $${\bf p}=(+4,+2,-2,-4,+4,-4).$$
\end{enumerate}
Then the CPB transmitted/stored sequence is $${\bf c}=(+4,+2,-2,-4,+4,-4,+2,+2,0,-4,-2,-2,+4).$$
\end{example}

\subsection{Symbol-Balanced Code Construction} \label{subsecKnutSB}

\noindent
At first sight, the Knuth approach may seem to be less suitable for generating $q$-ary SB sequences than for CB and PB sequences. Still, Mascella and Tallini presented Knuth-like SB construction methods which are based on maps exchanging alphabet symbols \cite{MT2005},\cite{MT2006}. By applying $q-1$ such maps, each guaranteeing that a particular symbol appears the desired number of times, symbol balancing is achieved. Here, we present another Knuth-like SB method which is similar to this Mascella-Tallini approach in the sense that it also operates in $q-1$ rounds, but is different in the sense that it adds in each round an appropriate balancing sequence to the data sequence, rather than performing specific symbol exchanges. Hence, our method is more in the spirit of the constructions presented in the previous subsections.

In order to encode a data sequence $\bf u$ of length $k=qm$ into an SB sequence $\bf x$, we propose the following Knuth-like algorithm. It consists of $q-1$ rounds, numbered $1$, $2$, $\ldots$, $q-1$, in which we will perform simple reversible manipulations on the data sequence, such that the end result is SB. In round $v$, we force there to be exactly $m=k/q$ symbols $-q+1+2v$ in the sequence, a condition that will not change anymore in the next rounds.  For $v=1,2,\ldots,q-1$, let
\begin{equation*}
{\cal A}_q^v=\{-q-1+2v,-q+1+2v,\ldots,q-1\},
\end{equation*}
i.e., ${\cal A}_q^v$ is the sub-alphabet consisting of the $q+1-v$ largest elements of the alphabet ${\cal A}_q$,
\begin{equation} \label{defM}
M_v({\bf y})=\max\{j\in{\cal A}_q^v: S_j({\bf y})\ge S_i({\bf y}) \mbox{ $\forall i\in{\cal A}_q^v$}\},
\end{equation}
and
\begin{equation} \label{defm}
m_v({\bf y})=\min\{j\in{\cal A}_q^v: S_j({\bf y})\le S_i({\bf y}) \mbox{ $\forall i\in{\cal A}_q^v$}\},
\end{equation}
where $\bf y$ is a sequence over the alphabet ${\cal A}_q$. Note that, for all $v$, $M_v({\bf y})$ is a symbol from ${\cal A}_q^v$ appearing most frequently in $\bf y$, while  $m_v({\bf y})$ is a symbols from ${\cal A}_q^v$ appearing least frequently in $\bf y$.

The algorithm is described as follows.

\noindent{\bf SB Encoding Procedure}
\begin{enumerate}
\item Set ${\bf u}_0={\bf u}$ and $v=1$.
\item Set $m_v=m_v({\bf u}_{v-1})$, $M_v=M_v({\bf u}_{v-1})$, and create ${\bf u}_v$ from ${\bf u}_{v-1}=(h_1,h_2,\ldots,h_k)$ by leaving all $h_i\notin{\cal A}_q^v$ unchanged and adding the value
\begin{equation} \label{mM}
\left\{\begin{array}{ll}
-q-1+2v-m_v & \mbox{ if $i\le i_v$,} \\
-q-1+2v-M_v & \mbox{ if $i> i_v$,} \\
\end{array}\right.
\end{equation}
to the $h_i\in{\cal A}_q^v$. The addition is done modulo $2q+2-2v$ such that the resulting symbol is in ${\cal A}_q^v$.
The value $i_v\in\{0,1,\ldots,k\}$ is chosen such that
\begin{equation} \label{symbal}
S_{-q-1+2v}({\bf u}_v)=k/q=m.
\end{equation}
\item If $v<q-1$, then increase $v$ by one and go back to the previous step.
\item Set ${\bf x}={\bf u}_{q-1}$, which is SB, and map $$(i_1,\ldots,i_{q-1},m_1,\ldots,m_{q-1},M_1,\ldots,M_{q-1})$$ to a unique SB prefix $\bf p$.
\end{enumerate}

\noindent Then transmit or store the SB codeword ${\bf c}=({\bf p},{\bf x})$.

\noindent{\bf SB Decoding Procedure}
\begin{enumerate}
\item  Retrieve $$(i_1,\ldots,i_{q-1},m_1,\ldots,m_{q-1},M_1,\ldots,M_{q-1})$$ from $\bf p$ and set ${\bf x}_{q}={\bf x}$ and $v=q-1$.
\item Create ${\bf x}_v$ from ${\bf x}_{v+1}=(h_1,h_2,\ldots,h_k)$ by leaving all $h_i\notin{\cal A}_q^v$ unchanged, and subtracting the value as given in (\ref{mM}) from the $h_i\in{\cal A}_q^v$. The subtraction is done modulo $2q+2-2v$ such that the resulting symbol is in ${\cal A}_q^v$.
\item If $v>1$, then decrease $v$ by one and go back to the previous step.
\item Set ${\bf u}={\bf x}_1$.
\end{enumerate}
\beginofproof
By construction we have $$S_{-q-1+2v}({\bf u}_w)=S_{-q-1+2v}({\bf u}_v)$$ for all $1\le v<w\le q-1$, and thus it follows from (\ref{symbal}) that all symbols from ${\cal A}_q$ appear equally often in ${\bf x}={\bf u}_{q-1}$, and thus $\bf x$ is SB. Further, it is easy to see that the operations in the encoding procedure are properly reversed in the decoding procedure. Hence, the only thing left to show is that for all $v=1,2,\ldots,q-1$ there always exists at least one $i_v$ such that (\ref{symbal}) is satisfied. From (\ref{defM}) and  (\ref{defm}), it follows that $S_{m_1}({\bf u}_0)\le m\le S_{M_1}({\bf u}_0)$, and thus
\begin{equation*}
S_{-q+1}({\bf u}_1)=S_{M_1}({\bf u}_0)\ge m \mbox{ if $i_1=0$,}
\end{equation*}
while
\begin{equation*}
S_{-q+1}({\bf u}_1)=S_{m_1}({\bf u}_0)\le m \mbox{ if $i_1=k$.}
\end{equation*}
Since increasing or decreasing $i_1$ by $1$ increases $S_{-q+1}({\bf u}_1)$ by $-1$, $0$, or $+1$, we can conclude that $S_{-q+1}({\bf u}_1)=m$ for at least one $i_1\in\{0,1,\ldots,k\}$. Similarly, for $v>1$, we have
\begin{equation*}
S_{m_v}({\bf u}_{v-1})\le \frac{k-(v-1)m}{q+1-v}= m\le S_{M_v}({\bf u}_{v-1}),
\end{equation*}
and thus
\begin{equation*}
S_{-q-1+2v}({\bf u}_v)=S_{M_v}({\bf u}_{v-1})\ge m \mbox{ if $i_v=0$,}
\end{equation*}
while
\begin{equation*}
S_{-q-1+2v}({\bf u}_v)=S_{m_v}({\bf u}_{v-1})\le m \mbox{ if $i_v=k$,}
\end{equation*}
and so $S_{-q-1+2v}({\bf u}_v)=m$ for at least one value $i_v\in\{0,1,\ldots,k\}$.
\endofproof

Note that there are at most $(k+1)^{q-1}$ possible realizations of $(i_1,\ldots,i_{q-1})$, $q!$ possible realizations of $(m_1,\ldots,m_{q-1})$, and $q!$ possible realizations of $(M_1,\ldots,M_{q-1})$. Hence, an unbalanced redundancy of $$p'= (q-1)\log_q (k+1)+2\log_q(q!)$$ suffices.
We conclude that, as for the Mascella-Tallini constructions \cite{MT2005}, \cite{MT2006}, the redundancy of this Knuth-like SB method exceeds the minimum redundancy by a factor of two for long codes.

\begin{example} \label{exSB}
Let $q=3$ and $n=6$, and thus the symbol frequency should be $m=6/3=2$. The data sequence is given to be $${\bf u}={\bf u}_0=(0,-2,-2,-2,0,-2).$$
Hence, $S_{-2}({\bf u}_0)=4$, $S_0({\bf u}_0)=2$, $S_{+2}({\bf u}_0)=0$, and thus $M_1=-2$ (the most frequent symbol) and $m_1=+2$ (the least frequent symbol). According to (\ref{mM}), in the first round ($v=1$), the number of $-2$ symbols is forced to be $2$ by modulo-6 adding $-4$ to the first $i_1$ symbols of ${\bf u}_0$ and $0$ to the last $6-i_1$ symbols. Choosing $i_1=3$ gives
$${\bf u}_1=(+2,0,0,-2,0,-2).$$
Note that $S_{-2}({\bf u}_1)=2$, $S_0({\bf u}_1)=3$, $S_{+2}({\bf u}_1)=1$, and thus $M_2=0$ and $m_2=+2$.
In the next round  ($v=2$), the number of zeroes is forced to be 2 by modulo-4
adding $-2$ to the first $i_2$ symbols  of ${\bf u}_1$ and $0$ to the last $6-i_2$ symbols, except when the symbol is equal to $-2$, in which case we leave it unchanged. Choosing $i_2=3$ gives
$${\bf u}_2=(0,+2,+2,-2,0,-2).$$
Note that $S_0({\bf u}_2)=S_1({\bf u}_2)=S_2({\bf u}_2)=2$, and thus ${\bf x}={\bf u}_2$ is SB.
\end{example}

\subsection{Discussion} \label{discKnuth}

\noindent
In the previous subsections, we have presented generalizations of Knuth's binary/bipolar balancing algorithm to larger alphabets, for the various balancing perspectives under consideration in this paper. Examples have been provided to illustrate the (encoding) procedures. It should be mentioned that these examples are misleading in the sense that the redundancy appears to be relatively large, which is due to the fact that extremely short data blocks were used in the examples. For instance, in Example~\ref{exPB}, four redundant symbols are used for eight data symbols. However, for long codes, the redundancy is only logarithmic in the length of the data block. For all the constructions presented in this section, the redundancy is roughly twice the corresponding minimum redundancy derived in Section~\ref{redundancy}.

For the binary case, modifications of Knuth's method have been presented to close the factor of two gap between the redundancy of the original Knuth algorithm and the minimum redundancy, while maintaining sufficient simplicity to enable feasible implementations. In \cite{IW2010}, this is done by a more efficient (variable-length) encoding of the prefix. In \cite{WI2010}, minimum redundancy is achieved by exploiting the fact that many data sequences have more than one possible balancing index, thus allowing to encode auxiliary data through the choice of the index. It is an interesting research challenge to investigate whether such techniques are also applicable in non-binary cases.

\section{Conclusions} \label{conc}

\noindent
In this paper we have considered balancing of $q$-ary sequences from various perspectives. In particular, we have reviewed the symbol balancing and charge balancing concepts, and introduced the polarity balancing concept, also in combination with charge balancing. For each of these perspectives, we have derived (approximate) expressions for the number of such sequences of a fixed length and for the minimum redundancy. The major conclusions of this analysis have been summarized in Table~\ref{normred}, which shows the minimum redundancy normalized to the logarithm of the block length $n$ in the limit as $n\rightarrow\infty$. Furthermore, we have presented for each of the balancing perspectives a $q$-ary coding scheme in the spirit of the binary Knuth algorithm. These schemes allow for simple encoding and decoding, at the price of a redundancy which is twice the minimum required redundancy.
%Closing this gap while maintaining sufficient simplicity, as already established for the binary case, is an interesting research challenge for $q$-ary %balanced codes.

\end{document}